\documentclass{elsart3}
\usepackage{natbib,epsfig}
\begin{document}
\runauthor{Marco Battaglia}
\begin{frontmatter}
\title{A Monolithic Pixel Sensor\\ in 0.15~$\mu$m Fully Depleted SOI Technology}
\author[UCB,LBNL]{Marco Battaglia,}
\author[INFN]{Dario Bisello,}
\author[LBNL]{Devis Contarato,}
\author[LBNL]{Peter Denes,}
\author[LBNL,INFN]{Piero Giubilato,}
\author[UCB]{Lindsay Glesener,}
\author[LBNL]{Chinh Vu}
\address[UCB]{Department of Physics, University of California at 
Berkeley, CA 94720, USA}
\address[LBNL]{Lawrence Berkeley National Laboratory, 
Berkeley, CA 94720, USA}
\address[INFN]{Dipartimento di Fisica, Universita' di Padova and INFN,
Sezione di Padova, I-35131 Padova, Italy}

\begin{abstract}
This letter presents the design of a monolithic pixel sensor with 10$\times$10~$\mu$m$^2$ 
pixels in OKI 0.15~$\mu$m fully depleted SOI technology and first results of its 
characterisation. The response of the chip to charged particles has been 
studied on the 1.35~GeV $e^-$ beam at the LBNL ALS. 
\end{abstract}
\begin{keyword}
Monolithic pixel sensor; SOI; CMOS technology; Particle detection
\end{keyword}
\end{frontmatter}

\typeout{SET RUN AUTHOR to \@runauthor}

Silicon on insulator (SOI) technology allows to fabricate CMOS circuits on a 
thin Si layer, electronically insulated from the rest of the wafer.
The isolation of the electronics from the detector volume offers clear 
advantages for designing monolithic pixel sensors for particle detection, compared 
to MAPS pixel devices, realised in standard CMOS bulk process. First both nMOS and 
pMOS transistors can be built, without disturbing the charge collection, and then the 
detector wafer can be biased thus improving the efficiency of charge carriers collection. 
There have been already attempts of developing SOI pixel sensors for charged particle 
detection with a high resistivity bottom wafer. 
The first used a 3~$\mu$m process at IET, Poland and gave a proof of principle 
of the concept~\cite{soi-iet1}; a test structure with 150$\times$150~$\mu$m$^2$
pixels observed signals of low-momentum electrons from a $^{90}$Sr source~\cite{soi-iet3}.
The availability of the 0.15~$\mu$m FD-SOI process by OKI Electric Industry Co.\ Ltd.\, 
Japan, has opened up new possibilities for SOI pixel sensors with the pixel pitch 
required for the next generation of particle physics experiments and imaging.
A chip based on this process has already been designed by a group at KEK, Japan and 
successful tested with an IR laser beam~\cite{soi-oki2}.

We designed and submitted a monolithic pixel sensor chip, with 10$\times$10~$\mu$m$^2$ 
pixels, for detection of charged particles. This letter presents the chip design, results 
of its characterisation and the first signal of high momentum particles, obtained with 
the 1.35~GeV electrons beam from the LBNL Advanced Light Source (ALS) booster. 

\begin{figure}[h!]
\begin{center}
\epsfig{file=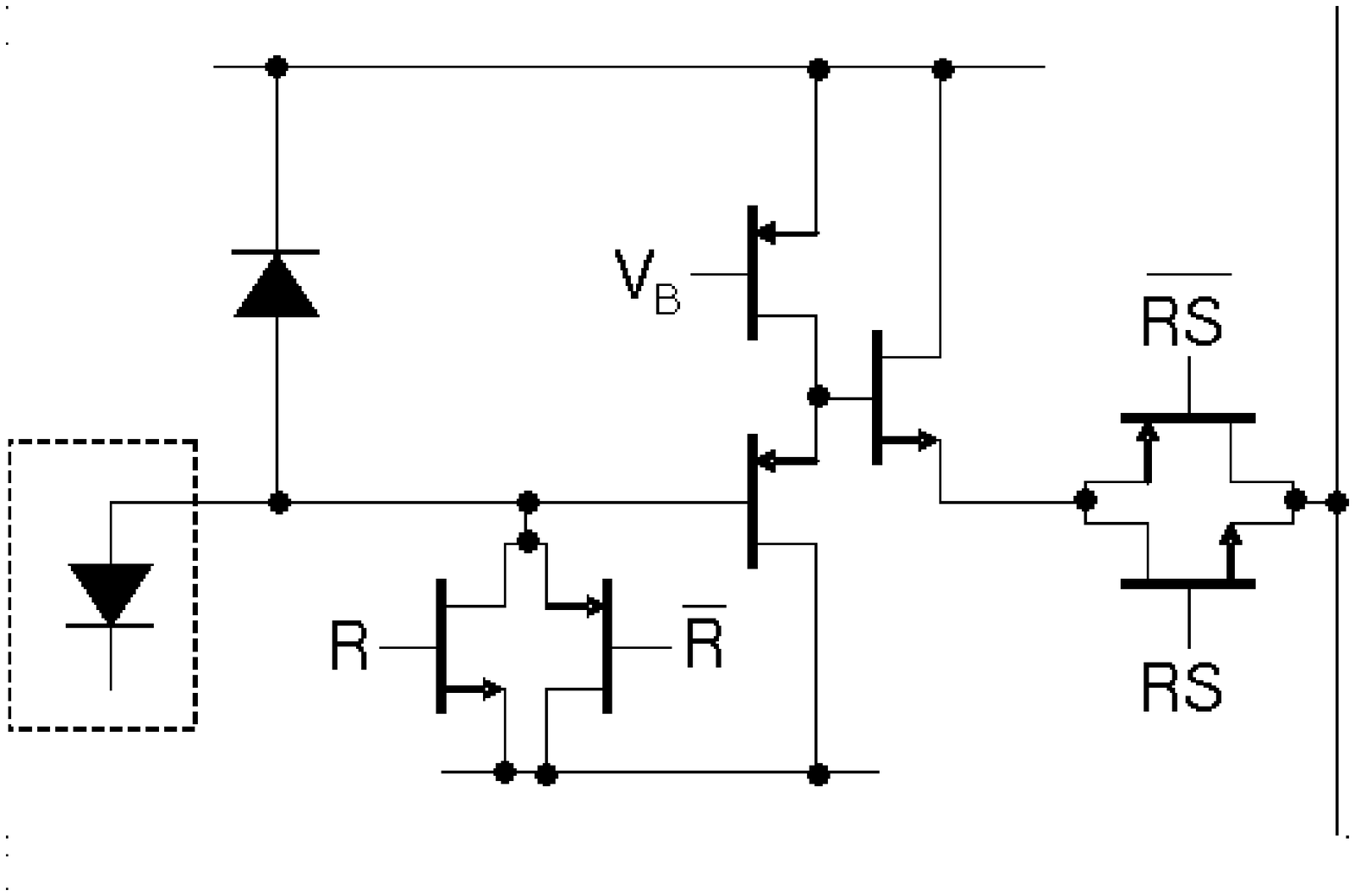,width=6.5cm}
\end{center}
\caption[]{Schematics of the analog pixel cell.}
\label{fig:schematics}
\end{figure}

The sensor has a 350~$\mu$m thick high-resistivity substrate, the CMOS circuitry 
is implanted on a 40~nm Si layer on top of  a 200~nm thick buried oxide. 
\begin{figure}[h!]
\begin{center}
\epsfig{file=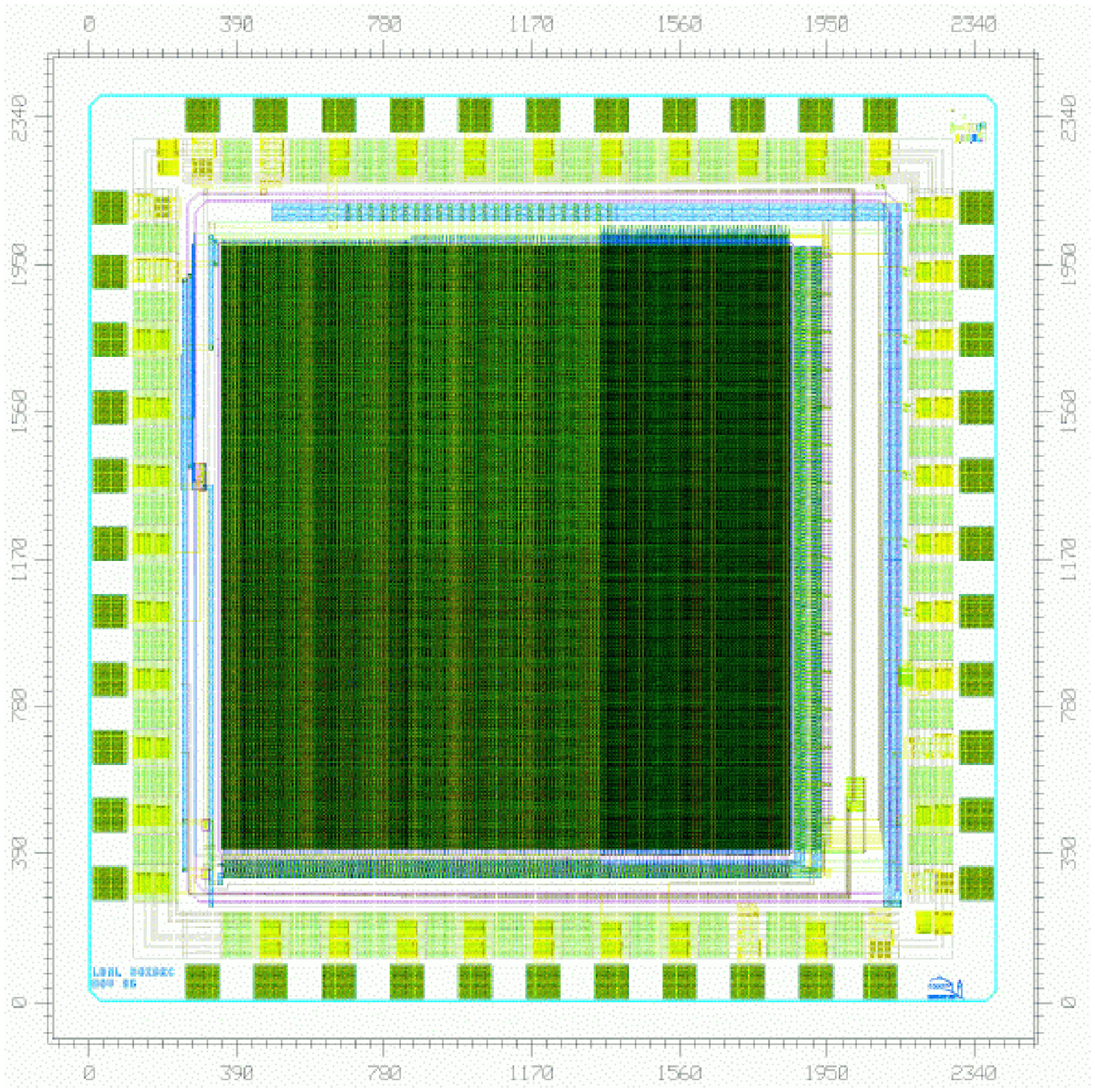,width=6.5cm}
\end{center}
\caption[]{Layout of the SOI pixel chip.}
\label{fig:layout}
\end{figure}
The thickness of the CMOS layer is small enough for the layer to be fully
depleted. The chip features an array of 160$\times$150 pixels on a 10~$\mu$m pitch.
The OKI SOI technology includes thin-oxide 1.0~V transistors and 
thick-oxide 1.8~V transistors.  The left-most 50 columns are simple analog pixels 
constructed with 1.0~V transistors.  The centre 50 columns are constructed with 
1.8~V pixels, and the right-most 50 columns are clocked, digital pixels.  The 
1.0~V pixels have significantly higher leakage currents than the 1.8~V pixels, 
which will be seen later to affect the signal-to-noise ratio.  
The schematic for the analog pixels is shown in Figure~\ref{fig:schematics}.  
The operation is comparable to a conventional 
3T pixel, except that an internal source follower has been added in order to minimise 
kickback from the row select switches. 
Each sector is divided in two subsections 
with 1$\times$1 and 5$\times$5~$\mu$m$^2$ charge collecting diodes. 
Single transistor test structures have been implemented at the chip periphery,
including complementary $p$-type and $n$-type MOSFETs, all with $W$=50~$\mu$m
and $L$=0.3~$\mu$m, with different types of body contacts (floating,
source-tied and gate-tied).
\begin{figure}
\begin{center}
\epsfig{file=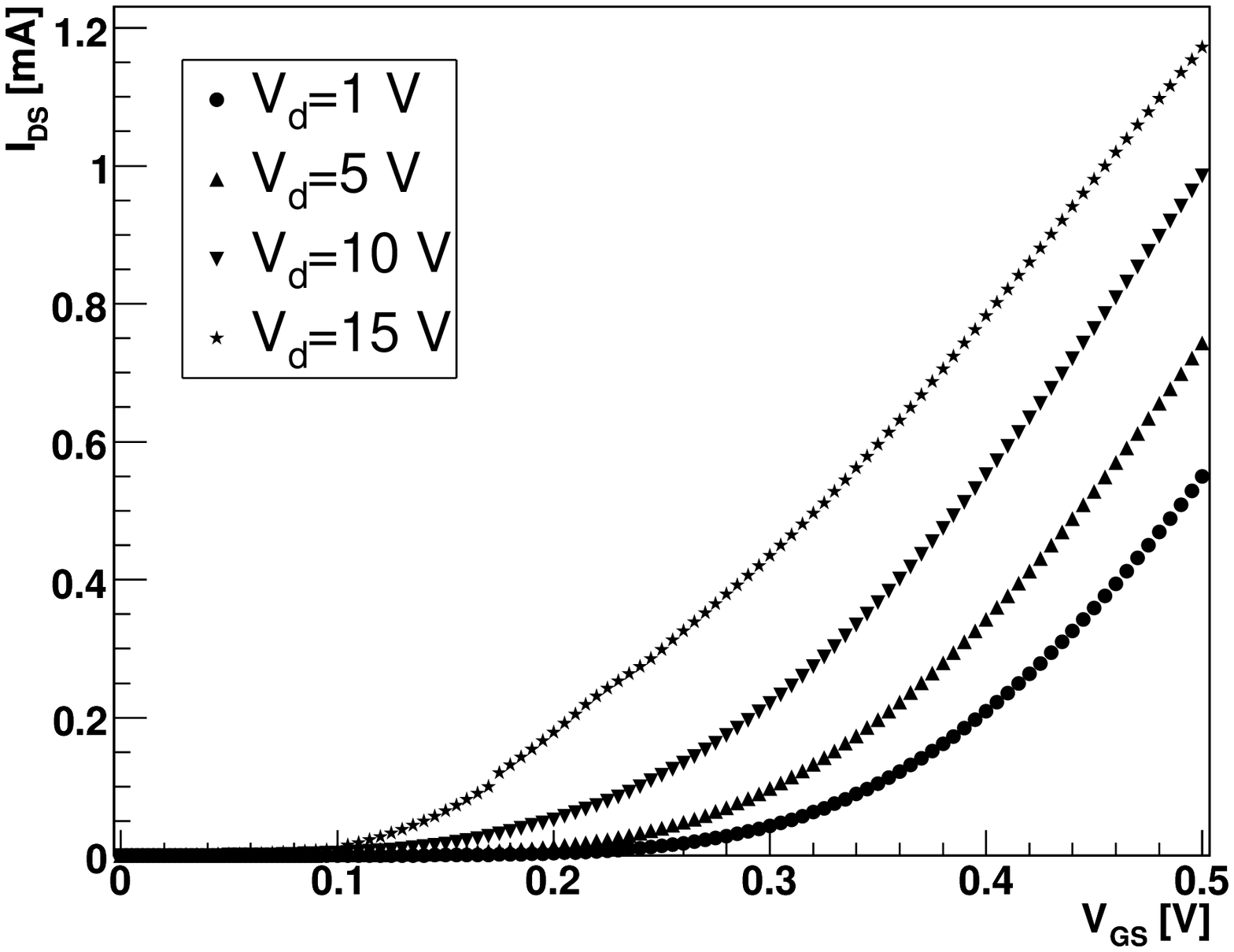,width=6.5cm}
\end{center}
\caption[]{Shift of input characteristics vs.\ substrate 
bias $V_d$ for 1.0~V $n$-MOS test transistor with $W/L$=50/0.3.}
\label{fig:threshold_shift}
\end{figure}
The transistor characteristics have been studied as a function of the sensor
substrate bias, $V_d$, in order to evaluate possible back-gating effects, which 
are expected to be significant due to the relatively small thickness of
the buried oxide. Figure~\ref{fig:threshold_shift} shows the input
characteristics $I_{DS}(V_{GS})$ measured at different substrate biases for a
nMOSFET with $W$=50~$\mu$m and $L$=0.3~$\mu$m, with floating body contact.
The measurements are performed with the transistor biased in saturation
region. The threshold voltages
shift from 0.24~V at $V_d$=1~V to 0.07~V at $V_d$=15~V, consistent with 
an increased back-gating effect.

Each 8000-pixel analog section is read out independently using a 14-bit ADC.  
A Xilinx FPGA controls all pixel clocks and resets.  The pixels are clocked at 
6.25~MHz giving an integration time of 1.382~ms. 
Correlated double sampling is performed by acquiring two frames of 
data with no pixel reset between the readings and subtracting the first frame 
from the second.  The response of the analog sections has been tested with an 
1060~nm IR laser, for different $V_d$ values.  The laser is focused to a 
$\simeq$20~$\mu$m spot and pulsed for 30~$\mu$s between successive readings.
We measure the signal pulse height in a 5$\times$5 matrix,  centred around the 
laser spot centre. The measured signal increases as 
$\sqrt{V_d}$, as expected from the increase of the depletion region, until 
$V_d \simeq$ 9~V, where it saturates, to decrease for $V_d \ge$ 15~V. 
\begin{figure}[h!]
\begin{center}
\epsfig{file=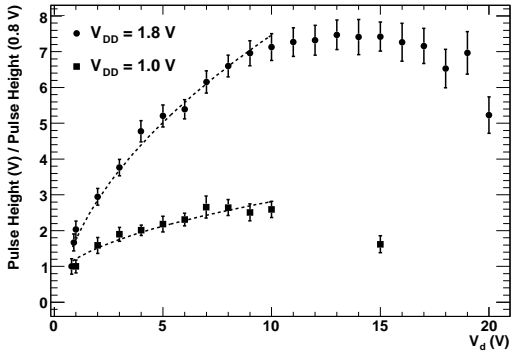,width=7.0cm}
\end{center}
\caption[]{Cluster pulse height normalised to that measured at 
$V_d$ = 0.8~V for a focused 1060~nm laser spot vs.\ $V_d$.}
\label{fig:laser}
\end{figure}
We interpret this effect as due to the transistor back-gating and affects the 
1.0~V transistor pixels at lower values of $V_d$.
The pixel chip has been tested on the 1.35~GeV electron beam-line at LBNL 
ALS. The readout sequence is synchronised 
with the 1~Hz booster extraction cycle so that the beam spill hits the 
detector just before the second frame is read. The temperature is kept 
constant during operation at $\simeq$~23$^{\circ}$C by forced airflow.
Only the analog part of the chip is readout. All the sections are 
functional and here we report results for two of the analog sections, with 
$V_{DD}$=1.0~V and 1.8~V transistors respectively, each consisting of a 
0.4$\times$0.8~mm$^2$ active region. 
Data are processed on-line by a LabView-based program, which performs 
correlated double sampling, pedestal subtraction and noise computation. 
The data is converted in the {\tt lcio} format and the 
offline analysis is performed using a set of dedicated processors developed in the 
{\tt Marlin} framework~\cite{Gaede:2006pj}. 
\vspace*{0.05cm}
\begin{figure}[h!]
\begin{center}
\epsfig{file=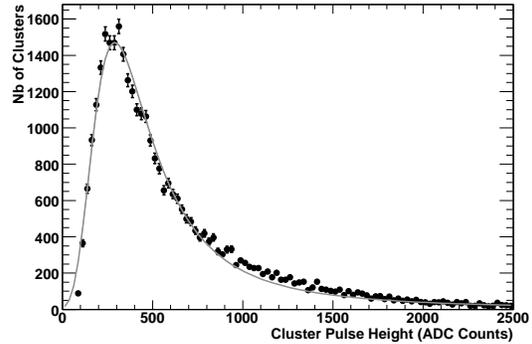,width=7.0cm}
\end{center}
\caption[]{Cluster pulse height distribution for 1.35~GeV $e^-$s for 
$V_d$ = 10~V, $V_{DD}$=1.8~V. The slight excess of 
events above to the fitted Landau function is interpreted as due to 
low momentum electrons in the beam.}
\label{fig:als}
\end{figure}
Each event is scanned for seed pixels
with pulse height above a signal-to-noise (S/N) threshold of 4.5. 
Noisy pixels are flagged and masked. 
Seeds are sorted according to their pulse heights and the 
neighbouring pixels in a 5$\times$5 matrix with S/N$>$2.5 
are added to the cluster. Clusters are not allowed to overlap 
and we require that pixels associated to a cluster are not interleaved by 
any pixels below the neighbour threshold. The average number of pixels 
accepted in a cluster is given in Table~\ref{tab:als}.
Only clusters with at least two pixels are further considered for the 
analysis. Data have been taken at different depletion voltages, 
$V_d$, from 1~V up to 20~V. Control data sets were taken under the same 
conditions but without beam, to monitor the effect of noisy pixels. 
Results are summarised in Table~\ref{tab:als}. There is only a small 
background arising from noisy pixels, which survives the bad pixel 
cut and the cluster quality criteria. Figure~\ref{fig:als} shows the cluster 
pulse height for data taken with $V_d$ = 10~V. The pixel 
multiplicity in a cluster decreases with increasing depletion voltage while the
cluster pulse height increases up to 10~V. At 15~V the cluster signal and the 
efficiency of the chip decreases, as observed in the laser test. 
\begin{table*}
\caption{Summary of ALS beam test results.
The average number of clusters per ALS spill recorded 
with beam on and beam off, average pixel multiplicity in a cluster inclusive of single 
pixel clusters, most probable value of cluster pulse height and signal-to-noise
ratio of seed pixels are given for different values of $V_d$. The beam intensity of the 
various runs was not constant.}
\label{tab:als}
\begin{center}
\begin{tabular}{|c|c|c|c|c|c|c|c|c|c|c|}
\hline  \textbf{$V_d$} & \multicolumn{2}{|c|}{\textbf{$\frac{Nb. Clusters}{Spill}$}} &  \multicolumn{2}{|c|}{\textbf{$\frac{Nb. Clusters}{Spill}$}} & \multicolumn{2}{|c|}{\textbf{$<$Nb Pixels$>$}} & \multicolumn{2}{|c|}{\textbf{Signal MPV}} & \multicolumn{2}{|c|}{\textbf{$<$S/N$>$}} \\ 
\textbf{(V)}    &  \multicolumn{2}{|c|}{\textbf{beam on}} & \multicolumn{2}{|c|}{\textbf{beam off}} & 
\multicolumn{2}{|c|}{\textbf{in Cluster}} & \multicolumn{2}{|c|}{\textbf{(ADC Counts)}}  & \multicolumn{2}{|c|}{}  \\ \hline
  & ~1.0V~ & ~1.8V~ & ~1.0V~ & ~1.8V~  & ~1.0V~ & ~1.8V~ & ~1.0V~ & ~1.8V~ & ~1.0V~ & ~1.8V~ \\ 
\hline
~1  & 3.9 & ~9.7 & 0.02 & 0.05 & 2.67 & 3.31 & 105. & 132 & ~7.4 & ~8.9 \\ 
~5  & 6.7 & 14.0 & 0.03 & 0.12 & 2.54 & 3.39 & 140. & 242 & ~8.8 & 14.9 \\
10  & 4.4 & ~7.8 & 0.03 & 0.20 & 2.41 & 3.31 & 164. & 316 & ~8.1 & 15.0 \\
15  & 1.4 & ~3.9 & 0.02 & 0.01 & 2.02 & 2.45 & 123. & 301 & ~6.5 & 13.6 \\ \hline
\end{tabular}
\end{center}
\end{table*}
The depletion voltages used correspond to an estimated depletion thickness 
from 8~$\mu$m to 56~$\mu$m for $V_d$ = 1~V to 15~V. The section with 1.8~V 
transistors exhibits a good signal-to noise ratio for 5~V$\le V_d \le$15~V.
These results are very encouraging for the further development of monolithic
pixel sensors in SOI technology.

\vspace*{-0.5cm}

\section*{Acknowledgements}

\vspace*{-0.25cm}

This work was supported by the Director, Office of Science, of the 
U.S. Department of Energy under Contract No.DE-AC02-05CH11231.
We are indebted to the ALS staff for their assistance and the 
excellent performance of the machine.

\end{document}